\documentclass[%
 reprint,
 amsmath,amssymb,
 aps,
]{revtex4-2}
\usepackage{physics}
\usepackage{graphicx}
\usepackage{dcolumn}
\usepackage{bm}
\usepackage{lineno}
\usepackage{adjustbox}
\usepackage{epsfig}
\usepackage{epstopdf}
\usepackage{float}
\usepackage{amssymb}
\usepackage{color}
\usepackage{bm}
\usepackage{amsfonts}
\usepackage{lineno,hyperref}
\usepackage{array}
\usepackage{microtype}
\usepackage{xcolor}
\usepackage{caption}
\usepackage{subcaption}

\begin{document}

\preprint{APS/123-QED}

\title{Stability and Thermodynamics of a Generalized Power-Law Dark Energy Model}
\author{S. Kazemi $^1$}%
\email{sepideh.kazemi5760@iau.ac.ir}

\author{M. A. Ramzanpour $^1,^2$}
\email{Corresponding Author: mohammad1352@iau.ac.ir}
\author{E. Yusofi $^2,^3$}
\email{eyusofi@ipm.ir}
\author{A. R. Amani $^1$}%
\email{al.amani@iau.ac.ir}
\affiliation{$^1$Department of Physics, AA.C., Islamic Azad University, Amol, Iran}

\affiliation{$^2$Innovation and Management Research Center, AA. C., Islamic Azad University, Amol, Iran}

\affiliation{$^3$School of Astronomy, Institute for Research in Fundamental Sciences(IPM), P. O. Box 19395-5531,Tehran, Iran}

\date{\today}

\begin{abstract}
We investigate a generalized power-law dark energy equation of state of the form \( p = w\rho - \beta\rho^m \) in a flat FLRW universe, analyzing its dynamical stability and thermodynamic consistency. The model exhibits a rich phase space structure, with an effective cosmological constant \(\rho^* = [(1+w)/\beta]^{1/(m-1)}\) emerging as a stable attractor for \((w < -1 ,~ m > 1)\). Notably, the universe evolves from an early de Sitter phase (\(w \to -1\)) to late-time de Sitter-like one with phantom crossing (\(w(z) < -1\)), aligning with DESI observations. Dynamical analysis reveals that the \(m > 1\) regime avoids ghost instabilities while accommodating phantom behavior, with \(m = 2\) providing particular theoretical advantages. Thermodynamically, the Generalized Second Law holds when the null energy condition \(\rho + p \geq 0\) is satisfied, which naturally occurs for \(\rho \geq \rho^*\). The model's compatibility with both observational data and fundamental thermodynamic principles suggests it as a viable framework for describing late-time cosmic acceleration, resolving tensions associated with phantom crossing while maintaining entropy dominance of the cosmological horizon.\\

\end{abstract}

\maketitle


\section{Introduction and Motivations}
\label{sec:intro}

The accelerated expansion of the universe, as evidenced by observations of Type Ia supernovae, cosmic microwave background (CMB) anisotropies, and large-scale structure surveys ~\cite{Riess1998, Perlmutter1999, Planck2018}, has prompted extensive investigations into the nature of dark energy (DE) . While the cosmological constant ($\Lambda$) remains the simplest explanation, it faces challenges such as the fine-tuning and coincidence problems ~\cite{Weinberg1989, Carroll2001}. Consequently, alternative models with dynamic equations of state (EoS) have been proposed to better capture the complexities of DE.

Traditional models often employ a linear EoS of the form $p = w \rho$, where $w$ is a constant. However, such models can lead to hydrodynamic instabilities, particularly when $w<-1$, corresponding to phantom energy scenarios~\cite{Caldwell2003}. To address these issues, generalized linear EoS models, such as affine parameterization of the dark sector $p =p_0 + \alpha \rho $~\cite{Pietrobon:2008js, Quercellini:2007ht}, have been introduced, allowing for a wider range of behaviors and improved stability characteristics. Further extensions involve non-linear EoS formulations~\cite{Guo2005, Ananda2006}, including quadratic terms~\cite{Ananda:2005xp, Mohammadi:2023idz, Moshafi:2024guo}, and Chaplygin gas~\cite{Bento2002} which can emerge from considerations in brane-world cosmologies and loop quantum gravity corrections has been been investigated in some studies~\cite{Li2007, Nojiri2003}.

The dynamical analysis of the stability of dark energy models is a crucial approach in cosmology to understand the long-term behavior and viability of proposed models describing the accelerated expansion of the universe. By examining the phase space of the underlying equations, cosmologists identify critical points and analyze their stability to determine whether the models can explain the observed cosmic acceleration consistently with a stable attractor solution. This test is essential because it helps distinguish physically plausible models from those that are dynamically unstable or lead to unrealistic future scenarios, thus providing valuable insights into the nature of dark energy and its underlying physics ~\cite{Guo2005,Alam2017,Wang2018,Cardenas2020}.

Beyond stability considerations, the thermodynamic consistency of DE models is evaluated through the Generalized Second Law (GSL) of thermodynamics~\cite{Camara2006, Wang2007}. The GSL posits that the total entropy, encompassing both the matter content and the horizon entropy, should not decrease over time \cite{Wang2016}. Studies have examined the validity of the GSL in various gravitational frameworks, including massive gravity theories like the Rham-Gabadadze-Tolley (RGT) model and its generalizations \cite{Kang2019}. These investigations often consider the entropy evolution within the apparent horizon and assess whether the GSL holds under different model parameters \cite{Bamba2017}.

In this study, a more realistic cosmic fluid is modeled as a multi-component system characterized by a self-merging power-law behavior proportional to $\rho^{m}$ \cite{Mohammadi:2023idz}. This system is described by a modified equation of state of the form
\begin{equation}\label{EQ:1}
	p = w \rho - \beta \rho^{m},
\end{equation}
where $w$, $\beta$, and $m$ are constant parameters governing the linear and nonlinear contributions to the pressure. The negative sign in the second term accounts for the negative pressure effects associated with dark energy. In this context, our aim is to investigate power-law extensions of the EoS model for dark energy, with a focus on their dynamical stability and adherence to the GSL. By integrating theoretical analyses with observational constraints, we endeavor to identify viable dark energy models that not only explain the observed accelerated expansion but also comply with fundamental physical principles.

The organization of this paper is as follows: Sec.  (\ref{sec:Model}), derives the basic equations within a flat FLRW universe. Sec. (\ref{ECC}), analyzes the effective cosmological constant and cosmic evolution. Sec. (\ref{sec:DyTest}) discusses the dynamical stability analysis, and Sec. (\ref{sec:GSLTest}) explores the GSL. Sec. \ref{sec:Desi} investigates the evolution of the dark energy EoS parameter $w(z)$, including phantom crossing. Finally, Sec. (\ref{sec:conclusions}) summarizes our main conclusions.

\section{Evolving Energy Density in the Non-Linear Equation of State Model}
\label{sec:Model}

The Einstein field equations are given by,
\begin{equation}
	R_{\mu\nu} - \frac{1}{2}g_{\mu\nu}R = T_{\mu\nu},
\end{equation}
where \( R_{\mu\nu} \) is the Ricci tensor, \( R \) is the Ricci scalar, \( g_{\mu\nu} \) is the metric tensor, and \( T_{\mu\nu} \) is the energy-momentum tensor of the cosmic fluid. We adopt natural units such that \( c^2 = 8\pi G = 1 \).

Assuming a homogeneous, isotropic, and spatially flat Friedmann-Lemaitre-Robertson-Walker (FLRW) spacetime, the metric reads,
\begin{equation}
	ds^2 = -dt^2 + a^2(t)\left[dr^2 + r^2(d\theta^2 + \sin^2\theta\,d\phi^2)\right],
\end{equation}
where \( a(t) \) is the scale factor, and \( (r, \theta, \phi) \) are the comoving coordinates. The FLRW metric is widely used in cosmology to describe the large-scale structure of the universe \cite{Weinberg1989, Carroll2001}.

Applying Einstein's equations to this metric and considering a perfect fluid with energy density \( \rho \) and pressure \( p \), we obtain the Friedmann equations,
\begin{align}
	3H^2 &= \rho, \label{eq:friedmann1} \\
	2\dot{H} + 3H^2 &= -p, \label{eq:friedmann2}
\end{align}
where \( H = \dot{a}/a \) is the Hubble parameter. These equations govern the dynamics of the universe and are fundamental to modern cosmology \cite{Riess1998, Perlmutter1999}.

The energy conservation equation for the cosmic fluid is,
\begin{equation}
	\dot{\rho} + 3H(\rho + p) = 0. \label{eq:conservation}
\end{equation}

We consider a non-linear dark energy equation of state of the form (\ref{EQ:1}). Such non-linear extensions of the equation of state have been explored in various cosmological contexts to address the limitations of linear models \cite{Nojiri2003, Bento2002}.

Substituting \eqref{EQ:1} into the conservation equation \eqref{eq:conservation}, we find the evolution of the energy density as a function of the scale factor:
\begin{equation}
	\rho(a) = \left[\frac{\beta}{1 + w} + C a^{-3(1 + w)(1 - m)}\right]^{\frac{1}{1 - m}}, \label{roa}
\end{equation}
where \( C \) is an integration constant. Notably, in the limit \( \beta \to 0 \), this reduces to the familiar form for a perfect fluid:
\begin{equation}
	\rho(a) = C a^{-3(1 + w)}. \label{eq:linear_limit}
\end{equation}

However, as \( w \to -1 \), the term \( \beta/(1 + w) \) becomes singular, indicating a more complex behavior near this limit that cannot be simply identified with a cosmological constant. This behavior has been extensively studied in the context of phantom energy and dark energy models \cite{Caldwell2003, Li2007}.


\section{Effective Cosmological Constant and Cosmic Evolution in Model}
\label{ECC}

The behavior of the cosmic evolution, effective cosmological constant (ECC), and regimes of \(\rho\) growth or decay depend on the parameters \(w\), \(\beta\), and \(m\) in the EoS given by Eq.~(\ref{EQ:1}). Such non-linear EoS models have been extensively studied to address the limitations of linear dark energy models and to better fit observational data \cite{Copeland2006, Tsujikawa2013}.

\subsection{Effective Cosmological Constant}
\label{subsec:ecc}

To find fixed points where \(\dot{\rho} = 0\) in the conservation equation, we solve:
\begin{equation}
	(1 + w)\rho - \beta\rho^m = 0,
\end{equation}
which yields for \(m \neq 1\),
\begin{equation}
	\rho^{m-1} = \frac{1 + w}{\beta}.
\end{equation}
The fixed point \(\rho^*\) is then:
\begin{equation}
	\rho^* = \left[\frac{1 + w}{\beta}\right]^{\frac{1}{m - 1}} \quad (w \neq -1, m \neq 1),
\end{equation}
assuming the expression inside the brackets is positive for real solutions. This \(\rho^*\) acts as an asymptotic or effective cosmological constant, a concept that has been explored in various modified gravity and dark energy scenarios \cite{Nojiri2017, Bamba2012}.

\subsection{Special Cases of the Model}
\label{subsec:special_cases}

\subsubsection{Case \(m = -1\)}
For \(m = -1\), the EoS takes the form:
\begin{equation}
	p = w\rho - \frac{\beta}{\rho}.
\end{equation}
The second term resembles the Chaplygin gas, which has been widely studied as a unified dark energy-dark matter model \cite{Bento2002, Kamenshchik2001}. The fixed points for the effective energy density \(\rho^*\) are:
\begin{equation}
	\rho^* = \pm\sqrt{\frac{\beta}{1 + w}}.
\end{equation}
Real solutions exist if \(\beta/(1 + w) \geq 0\). These solutions can act as attractors or repellers, corresponding to de Sitter-like regimes or other cosmological scenarios \cite{Bilic2002}.

\subsubsection{Case \(m = 0\)}
For \(m = 0\), the EoS reduces to:
\begin{equation}
	p = w\rho - \beta.
\end{equation}
The fixed point is:
\begin{equation}
	\rho^* = \frac{\beta}{1 + w},
\end{equation}
which corresponds to an affine equation of state. Such models have been used to describe dark energy with a constant offset and are consistent with certain observational constraints \cite{Quercellini2007, Pietrobon2008}.

\subsubsection{Case \(m = \frac{1}{2}\)}
For \(m = \frac{1}{2}\), the fixed point satisfies:
\begin{equation}
	\rho^* = \left(\frac{\beta}{1 + w}\right)^2.
\end{equation}
This case exhibits behavior akin to a transition between matter-like and dark energy-dominated phases, as discussed in the context of non-linear EoS models \cite{Ananda2006, Mukherjee2006}.

\subsubsection{Case \(m = 2\)}
For \(m = 2\), the fixed point is:
\begin{equation}
	\rho^* = \frac{1 + w}{\beta}.
\end{equation}
This quadratic case has been analyzed in detail and can model complex dynamical regimes, including phantom crossing and late-time acceleration \cite{Guo2005, Alam2017}.

\subsubsection{Case \(\beta = 0\)}
When \(\beta = 0\), the EoS simplifies to \(p = w\rho\), with the fixed point at \(\rho^* = 0\) for \(w \neq -1\). This corresponds to the standard linear fluid model, which has been extensively studied in the context of dark energy and cosmic acceleration \cite{Caldwell2002, Wang2018}. \textsf{Note that in all of these cases except for $m > 1$, ECC can only exist under condition \( w \neq -1 \) and it is singular in \( w = -1 \) .
} 
\begin{table*}[htbp]
	\centering
	\begin{tabular}{|c|c|c|c|c|}
		\hline
		\textsf{Special Cases} & \textsf{EoS} & \textsf{ECC} & \textsf{Comments}& \textsf{References} \\
		\hline
		\(\beta=0\) & \( p = w\rho \) & \( \rho^* = 0 \) & Standard linear fluid & Mostly\\
		\hline
		\(m=-1\) & \( p = w\rho - \beta \rho^{-1} \) & \( \rho^* = \pm \sqrt{\frac{\beta}{1 + w}} \) & Behaves as Chaplygin gas& \cite{Kamenshchik:2001cp, Bento2002} \\
		\hline
		\(m=0\) & \( p = w\rho - \beta \) & \( \rho^* = \frac{\beta}{1 + w} \) & Constant offset: linear EoS& \cite{Pietrobon:2008js, Quercellini:2007ht} \\
		\hline
		\(m=\frac{1}{2}\) & \( p = w\rho - \beta \sqrt{\rho} \) & \( \rho^* = \left(\frac{\beta}{1 + w}\right)^2 \) & Nonlinear term with fractional power& \cite{Roy:2024vhe} \\
		\hline
		\(m=2\) & \( p = w\rho - \beta \rho^2 \) & \( \rho^* = \frac{1 + w}{\beta} \) & Quadratic term& \cite{ Ananda:2005xp, Mohammadi:2023idz, Moshafi:2024guo} \\
		\hline
	\end{tabular}
	\caption{Summary of special cases for the generalized power-law EOS (\ref{EQ:1}).}
\end{table*}
\section{Dynamical Stability Analysis}
\label{sec:DyTest}

Understanding the dynamical behavior of cosmological models is essential for revealing their long-term evolution and stability. Such analysis provides valuable insights into the viability of models, their ability to describe the entire cosmic history, and their consistency with observational data. Consequently, dynamical studies play a crucial role in assessing the physical plausibility and predictive power of theoretical frameworks in cosmology~\cite{Copeland2006, Tsujikawa2013, Nojiri2017}. 

\subsection{Dynamical System Formulation}

The conservation equation for pressure in terms of \( N = \ln a \) in a flat Friedmann-Lemaître-Robertson-Walker (FLRW) universe is,
\begin{equation}
	\label{Dy1}
	\frac{d\rho}{dN} = -3 (1 + w) \rho + 3 \beta \rho^m,
\end{equation}

where \( N \) is the number of e-folds and \( a \) is the scale factor. To derive (\ref{Dy1}) we use
\[
\frac{d}{dt} = H \frac{d}{dN}.
\]. This form transforms the cosmological conservation equation into a first-order autonomous differential equation suitable for phase space analysis and stability study. It encapsulates the effects of the non-linear EoS parameters on cosmological dynamics, enabling investigation of fixed points, stability, and long-term behavior of the universe under this fluid model~\cite{Caldwell2002, Wang2016}.

\subsection{Critical Points}

Setting \(\frac{d\rho}{dN} = 0\) yields:
\begin{equation}
	0 = -3 (1 + w) \rho + 3 \beta \rho^m,
\end{equation}
which leads to critical points at:
\begin{itemize}
	\item \( \rho = 0 \) (the empty universe)
	\item \( \rho^* = \left[\frac{1 + w}{b}\right]^{\frac{1}{m-1}} \) (non-trivial solution)
\end{itemize}

\subsection{Stability Analysis of Fixed Points}
\label{subsec:stability}

Linear stability analysis reveals the behavior near each critical point. The eigenvalue for perturbations around a fixed point is:

\begin{equation}
	\lambda = \frac{d}{d\rho}\left(\frac{d\rho}{dN}\right) = -3(1 + w) + 3m\beta\rho^{m-1}.
	\label{eq:eigenvalue}
\end{equation}

\subsubsection{Vacuum Solution Stability}
For $\rho = 0$, the eigenvalue simplifies to:

\begin{equation}
	\lambda = -3(1 + w),
\end{equation}

leading to two distinct cases:
\begin{itemize}
	\item \textbf{Quintessence-like ($w > -1$)}: $\lambda < 0$ (stable attractor)
	\item \textbf{Phantom-like ($w < -1$)}: $\lambda > 0$ (unstable repeller)
\end{itemize}

This behavior matches known results for linear dark energy models \cite{Caldwell2002, Vikman2005}.

\subsubsection{Non-trivial Solution Stability}
For non-trivial vacuum-like $\rho = \rho^*$, the eigenvalue becomes,

\begin{equation}
	\lambda = 3(1 + w)(m - 1),
\end{equation}

with stability determined by,
\begin{equation}
	(1 + w)(m - 1) < 0 \quad \text{(Stability condition)}
\end{equation}

This leads to several physically interesting regimes:
\begin{itemize}
	\item $m > 1$ with $w < -1$: Stable phantom attractor
	\item $m < 1$ with $w > -1$: Stable quintessence attractor
	\item Other combinations: Unstable solutions
\end{itemize}

The stable phantom regime for $m > 1$ is particularly significant as it avoids the usual pathologies associated with phantom energy \cite{Nojiri2003, Carroll2004}.

\subsection{Physical Implications and Cosmic Evolution}

\begin{itemize}
	\item For \( m > 1 \):
	
	\textbf{(1)}\textsf{ When \( w < -1 \), the system in phantom phase approaches a stable de Sitter-like attractor characterized by a constant energy density.}
	
	\textbf{(2)} When \( w > -1 \), the system becomes unstable, with the pressure diverging over time.
	
	\item For \( m < 1 \):
	
	\textbf{(3)} When \( w > -1 \), the behavior is stable and resembles a fluid with pressure evolving as \( p \propto a^{-3(1 + w)} \).
	
	\textbf{(4)} When \( w < -1 \), the system exhibits instability.

The stability of the \(m\)EoS model in the phantom regime (\(w < -1\)) for \(m > 1\) is a significant advantage because it alleviates the common issue of ghost instabilities associated with phantom theories, enabling a theoretically consistent passage through the phantom divide. Additionally, this stability makes the model more compatible with observational hints of \(w < -1\), providing a better fit to current data. The ability to accommodate such observational preferences while maintaining internal consistency offers a more flexible and realistic description of cosmic acceleration, overcoming the stability challenges that typically hinder phantom models.
\end{itemize}
For $w \to -1$, singular behaviors appear in the solution.
This critical limit highlights the model's sensitivity to the phantom divide ($w=-1$), where for the generalized $m$ ECC becomes ill-defined. However, for $m>1$, the system naturally regularizes phantom behavior through the stable attractor $\rho^*$.

It is important to note that the case \(w = -1\), corresponding to a cosmological constant or de Sitter solution, is a special fixed point in the dynamical system. The stability analysis performed for \(w \neq -1\) relies on linearization around fixed points with well-defined eigenvalues, which may not hold at \(w = -1\) due to potential degeneracies or zero eigenvalues. In fact, de Sitter solutions are often known to be stable attractors in a wide range of dark energy models \cite{Copeland2006}. A more subtle analysis, such as center manifold techniques or higher-order perturbations, is typically required to rigorously determine the stability properties in this particular case. 
\subsection{Discussions on Different Regimes}
\label{subsec:regimes}

The evolution of the universe in this model depends critically on the ratio of \(\rho\) to the fixed point \(\rho^*\), which often functions as an effective cosmological constant or de Sitter solution. The following regimes are of particular interest:

\begin{itemize}
	\item \textbf{Regime A (Equilibrium)} \(\rho \approx \rho^*\): The universe approaches a de Sitter-like phase with near-constant energy density, driving late-time acceleration. Stability analysis (Sec.~\ref{sec:dynamical_stability}) shows this acts as an attractor for \(m > 1\), consistent with observations of cosmic acceleration \cite{Planck2020, DESI2024}.
	
	\item \textbf{Regime B (High-Density)} \(\rho > \rho^*\): For stable \(\rho^*\) (\(m > 1, w < -1\)), the energy density decays toward \(\rho^*\), naturally transitioning from early inflation to dark energy domination. For unstable configurations, \(\rho\) grows without bound, potentially leading to a Big Rip singularity, a phenomenon explored in phantom energy models \cite{Caldwell2003, Nojiri2005}.
	
	\item \textbf{Regime C (Low-Density)} \(\rho < \rho^*\): The system evolves toward \(\rho^*\) when stable (\(m < 1, w > -1\)), or away from equilibrium toward matter-dominated expansion when unstable. Such behavior has been observed in dynamical dark energy models and is compatible with large-scale structure surveys \cite{Eisenstein2005, Alam2017}.
\end{itemize}

\section{GSL test of the model}
\label{sec:GSLTest}
The GSL in thermodynamics plays a pivotal role in cosmology, particularly in assessing the physical viability of models \cite{Bousso2002, Wang2006, Cai2005}. According to the GSL, the sum of the entropy of matter enclosed within the cosmological horizon and the entropy of the horizon itself should not decrease over time \cite{Bak2000}. 
The validity of GSL in cosmology states that the total entropy, comprising the entropy of the apparent horizon $S_H$ and the entropy of the matter-energy contents inside it $S_m$~\cite{Jamil2014}, should never decrease over time. Mathematically:  
\begin{equation}  
	\frac{d}{dt} (S_H + S_m) \geq 0.  
\end{equation}  
In a flat FLRW universe, the radius of the apparent horizon is given by:  
\begin{equation}  
	r_A = \frac{1}{H},  
\end{equation}  
where $H$ is the Hubble parameter.  

Assuming the Bekenstein-Hawking entropy, the horizon entropy is proportional to its area:  
\begin{equation}  
	S_H = \frac{k_B c^3}{\hbar G} \pi r_A^2 \propto \frac{1}{H^2}.  
\end{equation}  
Adopting natural units ($k_B = c = \hbar = G = 1$), we write:  
\begin{equation}  
	S_H = \frac{\pi}{H^2}.  
\end{equation}  

The time derivative of $S_H$ is:  
\begin{equation}  
	\frac{d S_H}{dt} = -2 \pi \frac{\dot{H}}{H^3}.  
\end{equation}  

Using the Friedmann equation for a flat universe:  
\begin{equation}  
	\dot{H} = -\frac{1}{2} (\rho + p),  
\end{equation}  

we get:  
\begin{equation}  
	\frac{d S_H}{dt} = \pi \frac{\rho + p}{H^3}.  
\end{equation}  

The GSL condition becomes:  
\begin{equation}  
	\frac{d S_H}{dt} \geq 0 \quad \Rightarrow \quad \rho + p \ge 0,  
\end{equation}  
which coincides with the null energy condition (NEC).  

Early universe thermodynamics reveals a significant disparity between entropy contributions from matter and the cosmological horizon. As demonstrated in foundational studies \cite{KM:2023tyj, Egan_2010}, the matter entropy within the horizon becomes negligible compared to the horizon entropy itself. This hierarchy emerges because:

\begin{itemize}
	\item The matter entropy density either remains approximately constant or decays with cosmic expansion
	\item The horizon entropy, scaling as $S_H \propto H^{-2}$ with the apparent horizon area $A_h$, grows substantially during expansion
\end{itemize}

The resulting entropy budget becomes horizon-dominated, making GSL of thermodynamics effectively dependent on the horizon entropy evolution. This hierarchy simplifies the GSL validity condition to requiring NEC.
The dominance of horizon entropy persists across cosmic epochs where this energy condition holds.

Given the equation of state for the proposed model (\ref{EQ:1}), the NEC becomes,  
\begin{equation}  
	\rho + p = \rho (1 + w) - \beta \rho^m.  
\end{equation}  

Expressed in terms of $\rho$:  
\begin{equation}  
	\rho (1 + w) \geq \beta \rho^{m}.  
\end{equation}  

Dividing by $\rho^{m}$:  
\begin{equation}  
	\rho^{1 - m} (1 + w) \geq \beta.  
\end{equation}  

By rearranging we can obtain the following interesting result, 
\begin{equation}
	\label{con1}
	\rho \geq \rho^* = \left[  \frac{(1 + w)}{\beta} \right]^{\frac{1}{m - 1}}.
\end{equation}
Which is simultaneously compatible with regimes \textsf{A} and \textsf{B} defined for cosmological evolution in the previous section.
\section{Evolving Dark Energy, Phantom Crossing and Merging Voids}
\label{sec:Desi}

To analyze the evolution of the equation of state (EoS) parameter \(w(z)\) with redshift \(z\), the literature often adopts the Chevallier-Polarski-Linder (CPL) parametrization~\cite{Chevallier:2000qy, Linder:2002et},
\begin{equation}\label{EQ:CPL}
	w_{\rm CPL}(z) = w_0 + w_a \left(\frac{z}{1+z}\right),
\end{equation}
which captures the possible dynamical nature of dark energy. In our power-law \(m\)EoS model, utilizing equation (\ref{roa}), we derive an effective \(w(z)\) as a function of \(z\),
\begin{equation}\label{wzm}
	w(z) = w + \beta \left[\frac{\beta}{1 + w} + \frac{C}{(1+z)^{-3(1+w)(1 - m)}} \right]^{-1},
\end{equation}
highlighting the dynamical behavior of the EoS influenced by the evolving energy density and the model parameters. Note that in equation (\ref{wzm}), the variable $w$ does not represent the value at the present time, unlike in the CPL model.

As demonstrated in the previous section, the \(m\)EoS model with \(m > 1\) is more physically consistent, as it prevents singularities at \(w = -1\) via the critical energy density \(\rho^*\), and exhibits stable behavior even in the phantom regime (\(w < -1\)), aligning well with recent DESI DR2 results ~\cite{DESI:2025zgx, Giare:2024gpk}. In the physically motivated case of \(m=2\)\cite{Mohammadi:2023idz, Moshafi:2024guo}, equation (\ref{wzm}) reduces to
\begin{equation}\label{EQ:wm2}
	w(z) = w - \frac{w_a (1 + w)}{(1 + w - w_a)(1+z)^{-3(1+w)} + w_a}.
\end{equation}
To relate our parameters to the familiar CPL form, we identify \(w_a \equiv \beta \rho_0\). In the specific case where the dark energy component behaves as a cosmological constant (\(w = -1\)), equation (\ref{EQ:wm2}) reduces to\cite{Mohammadi:2023idz},
\begin{equation}
	\label{wdezsp}
	w(z) = -1 - \frac{w_a}{1 - 3 w_a \ln (1+z)},
\end{equation}
which provides a suitable description of a dynamical dark energy crossing the phantom divide. The variable values and signs of $w_a$ in the effective EoS \eqref{wdezsp} allow its parameter $w(z)$ to shift towards super negative~\cite{Caldwell2002} , zero, or even positive values, potentially aligning with recent DESI DR2 findings (see Fig. (12) in ~\cite{DESI:2025zgx}).

Also, it can be shown that, in the limits $z \rightarrow 0$ and $z \rightarrow \infty$, the dark energy equation of state parameter $w(z)$ yields the following interesting results,
\begin{equation}
	\label{slozz2}
	w(z\rightarrow \infty)\equiv w_{\rm inf}=-1
\end{equation}
\begin{equation}
	\label{slozz}
	w(z\rightarrow0)\equiv	w_{\rm de}=-1-w_a ~.
\end{equation}
Interestingly, these answers are consistent with condition (\ref{con1}). Our analysis of the dynamical and thermodynamic stability indicates that the universe initially undergoes an expansion from a cosmological constant (\(w_{\rm inf} = -1\)), corresponding to a pure de Sitter phase characterized by very high energy density at early times. As the universe evolves, entropy increases while the energy density decreases, ultimately approaching a non-zero effective cosmological constant phase with an EoS parameter \(w_{\rm de} =-1-w_a (\neq -1)\). Owing to the flexibility in the magnitude and sign of \(w_a\), the dark energy equation of state parameter in our model can be compatible with various observational datasets, including the Planck measurements as well as the DESI and WMAP data~\cite{Planck:2018jri, Escamilla:2023oce, Yusofi:2018lqb}. Overall, the \(m\)EoS offers enhanced flexibility and stability advantages, making it more suitable for modeling complex dark energy dynamics consistent with current observational constraints. 
\subsection{Merging Voids as a Possible Candidate for Cosmic Acceleration}  

The dual merging of cosmic voids introduces a second-order correction to the pressure of the cosmic fluid, as demonstrated in~\cite{Mohammadi:2023idz} and observationally tested in~\cite{Moshafi:2024guo}. This additional term modifies the effective EoS of the universe, potentially contributing to late-time acceleration. Intriguingly, our analysis shows that a second-order term can lead to stable dynamical and thermodynamic behavior, even in phantom regimes, while allowing for variable dark energy. This stability is crucial, as it avoids the pathological instabilities typically associated with phantom energy and aligns with observational constraints. The dynamical behavior arising from void mergers could thus provide a viable mechanism for the observed accelerated expansion without invoking exotic forms of dark energy.  

Recent observations from DESI DR2~\cite{DESI:2025zgx} further support the plausibility of this scenario, as they hint at dynamical dark energy behavior that may not be fully explained by a cosmological constant. The merging voids scenario naturally accommodates such dynamics, as the second-order pressure term evolves with cosmic structure formation. Unlike static dark energy models, this framework ties cosmic acceleration to the large-scale structure of the universe, offering a testable connection between void statistics and the expansion history~\cite{CosmoVerse:2025txj}. If confirmed, this would position merging voids as a compelling candidate for explaining dark energy, bridging the gap between structure formation and late-time cosmology~\cite{Benisty:2023vbz}. While the possibility of voids explaining the accelerating expansion without dark energy is intriguing, it still requires further investigation and validation through more comprehensive data and simulations. 

\section{Conclusions}
\label{sec:conclusions}
In this work, we investigated a power-law term for dark energy EoS and examined its dynamical and thermodynamic stability within a flat FLRW framework. Our analysis, particularly with recent DESI data, shows the model inherently accommodates the evolving EoS parameter \( w(z) \), transitioning from early de Sitter-like behavior (\( w \rightarrow -1 \)) to late-time quintessence with phantom crossing (\( w(z) < -1 \)). The stability analysis reveals that for \( m > 1 \), the universe tends toward stable de Sitter attractors, consistent with observations, while for \( m < 1 \), instabilities can occur depending on the parameter space. Additionally, the thermodynamic study confirms the validity of the GSL when the null energy condition holds. Overall, the flexible non-linear EoS model effectively captures the complex dark energy dynamics, including phantom crossing, while remaining compatible with current observational and fundamental physical laws, making it a promising candidate for describing the universe’s late-time acceleration. In general, our analysis of the proposed generalized power-law model shows that the model behaves very well with \(m > 1\), especially \(m=2\), to explain observations on the one hand and also has good compatibility with the fundamental principles of physics. In future work, we will focus on testing the parameters of this model using statistical methods and observational constraints.
\section*{Acknowledgement}
This work has been supported by the Islamic Azad University, Ayatollah Amoli Campus (AA.C.), Amol, Iran.
\section*{Data Availability}
There is no new data associated with this article.
\vspace{1cm}

\bibliographystyle{apsrev4-1}
\bibliography{Kazemi_p1}

\end{document}